\documentclass[prx,twocolumn,floatfix,superscriptaddress,amsmath,aps,longbibliography]{revtex4-1}
\pdfoutput=1
\usepackage{dcolumn,graphicx,color,booktabs,microtype,afterpage}
\usepackage[colorlinks,plainpages=false,linkcolor=blue,urlcolor=blue,citecolor=blue,pdfpagemode=UseNone,pdfstartview=FitBH]{hyperref}
\usepackage{graphicx}
\usepackage{bmpsize}
\usepackage{amsfonts}
\usepackage{gensymb}
\usepackage{amssymb}
\usepackage{bm}

\begin{document}
\title{Spin excitations in the kagome-lattice metallic antiferromagnet Fe$_{0.89}$Co$_{0.11}$Sn}
\author{Tao Xie{\color{blue}{$^{\S}$}}}
\thanks{Corresponding author: xiet@ornl.gov}
\thanks{$^{\S}$~These authors contributed equally to this work}
\affiliation{Neutron Scattering Division, Oak Ridge National Laboratory, Oak Ridge, Tennessee 37831, USA}
\author{Qiangwei Yin{\color{blue}{$^{\S}$}}}
%\thanks{These authors contributed equally to this work}
\affiliation{Laboratory for Neutron Scattering, and Beijing Key Laboratory of Optoelectronic Functional Materials MicroNano Devices, Department of Physics, Renmin University of China, Beijing 100872, China}
\author{Qi Wang}
\affiliation{Laboratory for Neutron Scattering, and Beijing Key Laboratory of Optoelectronic Functional Materials MicroNano Devices, Department of Physics, Renmin University of China, Beijing 100872, China}
\author{A. I. Kolesnikov}
\affiliation{Neutron Scattering Division, Oak Ridge National Laboratory, Oak Ridge, Tennessee 37831, USA}
\author{G. E. Granroth}
\affiliation{Neutron Scattering Division, Oak Ridge National Laboratory, Oak Ridge, Tennessee 37831, USA}
\author{D. L. Abernathy}
\affiliation{Neutron Scattering Division, Oak Ridge National Laboratory, Oak Ridge, Tennessee 37831, USA}
\author{Dongliang Gong}
\affiliation{Department of Physics and Astronomy, University of Tennessee, Knoxville, Tennessee 37996, USA}
\author{Zhiping Yin}
\affiliation{Center for Advanced Quantum Studies and Department of Physics, Beijing Normal University, Beijing 100875, China}
\author{Hechang Lei}
\thanks{Corresponding author: hlei@ruc.edu.cn}
\affiliation{Laboratory for Neutron Scattering, and Beijing Key Laboratory of Optoelectronic Functional Materials MicroNano Devices, Department of Physics, Renmin University of China, Beijing 100872, China}
\author{A. Podlesnyak}
\thanks{Corresponding author: podlesnyakaa@ornl.gov}
\affiliation{Neutron Scattering Division, Oak Ridge National Laboratory, Oak Ridge, Tennessee 37831, USA}

%\date{\today}

\begin{abstract}
Kagome-lattice materials have attracted tremendous interest due to the broad prospect for seeking  superconductivity, quantum spin liquid states,  and topological electronic structures. Among them, the transition-metal kagome lattices are high-profile objects for the combination of topological properties, rich magnetism, and multiple-orbital physics. Here we report an inelastic neutron scattering study on the spin dynamics of a kagome-lattice antiferromagnetic metal  Fe$_{0.89}$Co$_{0.11}$Sn. Although the magnetic excitations can be observed up to $\sim$250 meV, well-defined spin waves are only identified below $\sim$90 meV and can be modeled using Heisenberg exchange with ferromagnetic in-plane nearest-neighbor coupling $J_1$, in-plane next-nearest-neighbor coupling $J_2$, and antiferromagnetic (AFM) interlayer coupling $J_c$ under linear spin-wave theory. Above $\sim$90 meV, the spin waves enter the itinerant Stoner continuum and become highly damped particle-hole excitations. At the $K$ point of the Brillouin zone, we reveal a possible band crossing of the spin wave, which indicates a potential Dirac magnon. Our results uncover the evolution of the spin excitations from the planar AFM state to the axial AFM state in Fe$_{0.89}$Co$_{0.11}$Sn, solve the magnetic Hamiltonian for both states, and confirm the significant influence of the itinerant magnetism on the spin excitations.
\end{abstract}

\maketitle

\section{Introduction}
The magnetic kagome lattice, a two-dimensional (2D) network of corner-sharing triangles surrounding hexagons, provides an ideal platform to search for exotic states such as quantum spin liquids~\cite{Balents2010spin,Wen2019experimental,Broholm2020,lhotel2020fragmentation} and other topological quantum states ~\cite{nakatsuji2015largeAHE,Wang2018large,liu2018giantAHE,ye2018massive,Yin2018giant,Liu2020,Liu2021spin,Wang2021YMnSn,Kang20201,Yin2021probing}. Theoretical studies have shown that the typical electronic bands of kagome lattices contain linearly dispersive Dirac bands and nondispersive flat bands~\cite{Guo2009topological,mazin2014theoretical}. When a kagome lattice is occupied by 3$d$ transition-metal atoms, the combination of the rich magnetism, topological electronic bands, and multiple-orbital characteristics will induce abundant novel phenomena such as the anomalous Hall effect~\cite{nakatsuji2015largeAHE,ye2018massive,Yin2018giant,Wang2018large,liu2018giantAHE,Liu2021spin,Wang2021YMnSn}.

\begin{figure*}[t]
\center{\includegraphics[width=1\linewidth]{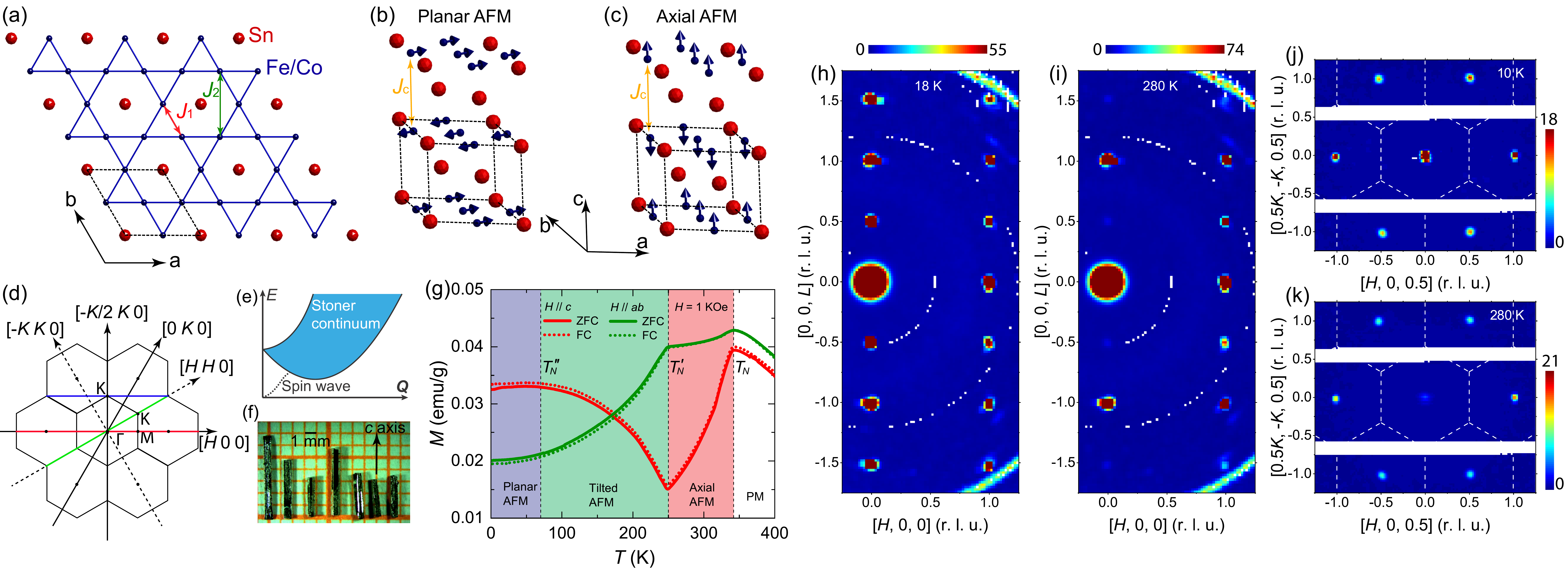}}
  \caption{~(a) The Fe-Co kagome layer with the  hexagonal holes filled with Sn in Fe$_{1-x}$Co$_x$Sn. The in-plane nearest-neighbor and next-nearest-neighbor exchange couplings are indicated by the red and green arrows, respectively. (b) and (c) The magnetic structures for the planar and axial AFM states in Fe$_{1-x}$Co$_x$Sn. The orange arrows represent the interlayer nearest-neighbor exchange coupling $J_c$. (d) A schematic of the 2D Brillouin zone of Fe$_{1-x}$Co$_x$Sn. (e) A schematic of the spin wave and Stoner continuum in some metallic magnets. (f) A typical photo of Fe$_{0.89}$Co$_{0.11}$Sn single crystals on 1-mm grid paper. The long axis of the rodlike crystals is the crystallographic $c$ axis. (g) The magnetization as a function of temperature of our Fe$_{0.89}$Co$_{0.11}$Sn sample. FC, field cooled; ZFC, zero-field cooled. (h) and (i) Zero-energy ($E$ = 0 $\pm$ 0.2 meV) 2D slices in ($H$ 0 $L$) with $E_i$ = 10 meV at 18 K (planar AFM state) and 280 K (axial AFM state), respectively. The arcs at the lower right corner and the upper right corner are the scattering intensity from the aluminum sampler holder. (j) and (k) Zero-energy ($E$ = 0 $\pm$ 1 meV) 2D slices in ($H$ $K$ 0.5) with $E_i$ = 80 meV at 10 K (planar AFM state) and 280 K (axial AFM state), respectively. The dashed lines indicate the boundary of the Brillouin zones in the ($H$ $K$ 0) plane.
  }
  \label{fig1}
\end{figure*}

In modern magnetic theory, the interaction between electron spins can be described by a local moment picture or itinerant electron model~\cite{Moriya1985,Romanov1988,melnikov2018,Coey2021handbook}. Although the former case is usually appropriate in the magnetic insulators and the latter model always comes into play in metallic magnets, some metallic systems can be analyzed by the local moment model~\cite{Ishikawa1078,Endoh2006}. What is more, both the local moment and the itinerant electron scenarios can coexist in some systems~\cite{zhao2009spin,Dai2015review,Jooseop2018Dual}. For example, in iron-based superconductors, the spin waves can be reproduced by an effective Heisenberg Hamiltonian taking into consideration the anisotropic spin-wave damping characteristics of an itinerant electron system~\cite{zhao2009spin,Dai2015review}. In some itinerant magnets, well-defined spin-wave excitations only can be observed in the low-energy region before entering the Stoner continuum [Fig.~\ref{fig1}(e)], in which the spin waves decay into damped particle-hole excitations~\cite{Mook1971spin,Mook1974magnetic,Ishikawa1078,Romanov1988,melnikov2018,Chen2020unconventional,Coey2021handbook}.

Recently, Dirac fermions and flat electronic bands have been reported in kagome-lattice metallic antiferromagnet FeSn, paramagnet CoSn, and the doped compounds Fe$_{1-x}$Co$_x$Sn by angle-resolved photoemission spectroscopy studies~\cite{Kang20201,Liu2020,Lin2020,Moore2022}. The family including FeSn and CoSn has a hexagonal structure with $P$6/$mmm$ space group. The Fe and Co atoms form the kagome lattice with hexagonal holes filled with Sn atoms [Fig.~\ref{fig1}(a)]. In antiferromagnetic (AFM) FeSn, below the N\'{e}el temperature $T_N$ = 365 K, the magnetic moments of Fe in each kagome layer align ferromagnetically, and the adjacent ferromagnetic (FM) layers stack in an antiparallel manner along the $c$ axis [Fig.~\ref{fig1}(b)]~\cite{Yamaguchi1967,Hggstrm1975,Kakihana2019FeSn,Sales,Lin2020,Kang20201}. With Co substitution at the Fe site in FeSn, the ordered moments' direction can be tuned from that in the $ab$ plane (planar AFM) [Fig.~\ref{fig1}(b)] to along the $c$ axis (axial AFM) [Fig.~\ref{fig1}(c)] continuously by crossing an intermediate state (tilted AFM)~\cite{Meier2019,Sales2021}. During this process, the magnetic moments of the neighboring FM layers remain antiparallel to each other. At some specific levels of Co doping, these different AFM states can be obtained by simply changing the temperature~\cite{Meier2019}.

Although the recent inelastic neutron scattering (INS) studies on FeSn show some differences, both reports confirm the non-negligible effect of itinerant electrons on the spin excitations~\cite{FeSn2021flat,do2021damped}, which suggests that the combination of localized and itinerant magnetism should be considered in this kind of metallic kagome-lattice AFM. While theoretical calculations suggested the existence of a magnetic flat band in FeSn~\cite{FeSn2021flat}, INS studies did not observe it~\cite{FeSn2021flat,do2021damped}. In addition, a damped Dirac magnon has been suggested to exist in FeSn~\cite{do2021damped}. The Co substitution in FeSn enriches the magnetism and may change the itinerancy of the electrons~\cite{Sales,Meier2019}, which makes Fe$_{1-x}$Co$_{x}$Sn a good candidate to study the topological magnon, the magnetic flat band, and their interplay with the itinerant electrons as well as the evolution of these properties with spin orientations.

In this paper, we select Fe$_{0.89}$Co$_{0.11}$Sn, which contains axial AFM, tilted AFM, and planar AFM states in different temperature regions, as our research object. By employing magnetization and neutron scattering measurements, we first confirm the existence of the different AFM states in Fe$_{0.89}$Co$_{0.11}$Sn. We subsequently obtain the in-plane FM spin excitations from zero energy to $\sim$250 meV together with the out-of-plane AFM spin wave below $\sim$25 meV in both planar and axial AFM states, which suggests quasi-2D magnetism in Fe$_{0.89}$Co$_{0.11}$Sn. The magnetic excitations below $\sim$80--90 meV can be described by linear spin-wave theory (LSWT) simulation. Above $\sim$90 meV, the spin waves enter the Stoner continuum and decay into the highly damped particle-hole excitations. Evidence of the existence of the Dirac magnon is also observed at the $K$ point of the Brillouin zone (BZ), albeit its upper part is obscured due to the interaction with the Stoner continuum from itinerant magnetism.

\section{EXPERIMENTAL DETAILS}
 We prepared high-quality single crystals of Fe$_{0.89}$Co$_{0.11}$Sn using the self-flux method. Details can be found in the Supplemental Material (SM)~\cite{SI} (see also Refs.~\cite{xu2013normalization,shirane2002neutron} therein). The crystals are long bars along the crystalline $c$ axis with a hexagonal cross section [Fig.~\ref{fig1}(f)]. Magnetization measurements were performed using a Quantum Design (QD) Magnetic Properties Measurement System (MPMS3). We coaligned about 100 single crystals in the ($H$ 0 $L$) scattering plane on thin aluminum plates to obtain a mosaic sample with a mass of about 2 g and mosaicity below 1.5$\degree$~\cite{SI}. The neutron scattering experiments were performed on the time-of-flight Wide Angular-Range Chopper Spectrometer (ARCS)~\cite{ARCS}, Fine-Resolution Fermi Chopper Spectrometer (SEQUOIA)~\cite{SEQ}, and Cold Neutron Chopper Spectrometer (CNCS)~\cite{CNCS1,CNCS2} at the Spallation Neutron Source, Oak Ridge National Laboratory. Measurements were carried out with a series of incident neutron energies $E_i$ = 3.32, 10, 80, 150, 250, 300, and 400 meV in both the planar AFM (at $T$ = 10 and 18 K) and axial AFM (at $T$ = 280 K) states. The sample was rotated along the vertical axis in a wide angle range (except for the $E_i$ = 400 meV measurement, where the beam was fixed with the $k_i$ $\parallel$ $c$ axis) to make a complete survey in the energy and momentum space. We used the software packages~\textsc{Mantid}~\cite{Mantid} and~\textsc{Horace}~\cite{Horace} for neutron scattering data reduction and analysis. The neutron scattering intensities are normalized to the same scale with arbitrary units using the incoherent elastic scatterings of the sample~\cite{SI}. Throughout this paper, a wave vector $\bf{Q}$ will be shown in reciprocal-lattice units (r. l. u.), in which $\bf{Q}$ = ($H$, $K$, $L$) means $\bf{Q}$ = $H$$\bf{a}^{*}$ + $K$$\bf{b}^{*}$ + $L$$\bf{c}^{*}$, where $\bf{a}^{*}$, $\bf{b}^{*}$, and $\bf{c}^{*}$ are basis vectors in reciprocal space.

\section{Neutron Scattering Results}
\subsection{Elastic neutron scattering}
The magnetization measurements of Fe$_{0.89}$Co$_{0.11}$Sn show three characteristic temperatures, named $T_N$~$\approx$~340 K, $T_{N}^{'}$~$\approx$~250 K, and $T_{N}^{''}$~$\approx$~70 K [Fig.~\ref{fig1}(g)], which correspond to the phase-transition temperatures from the paramagnetic (PM) state to the axial AFM state, the axial AFM state to the tilted AFM state, and the tilted AFM state to the planar AFM state, respectively~\cite{Meier2019}. In order to confirm the magnetic phases in our sample, we first check the elastic neutron scattering results and then compare the results with those in Ref.~\cite{Meier2019}.

Figures~\ref{fig1}(h)--\ref{fig1}(k) present several zero-energy 2D slices in the ($H$ 0 $L$) and ($H$ $K$ 0.5) planes. In Fig.~\ref{fig1}(h), we can see strong peaks appear at $Q$ = (0, 0, $n$/2) ($n$ = integer) and $Q$ = (1, 0, $n$/2). The peaks at integer $L$ are nuclear peaks, while the peaks at half-integer $L$ are magnetic Bragg peaks, which correspond to a propagation vector $\mathbf{q}$ = (0, 0, 1/2). The peaks in the ($H$ $K$ 0.5) plane shown in Fig.~\ref{fig1}(j) further confirm the fact that the ordered moments align ferromagnetically in the $ab$ plane. These observations are consistent with the previous neutron diffraction results~\cite{Meier2019}. Since neutron scattering measurements probe the magnetic moment components that are normal to the wave vector $\mathbf{Q}$, we expect no magnetic Bragg peaks at $Q$ = (0, 0, $n$ + 1/2) for an axial AFM state due to the parallel direction between magnetic moments and $\mathbf{Q}$'s [Fig.~\ref{fig1}(c)]. However, weak magnetic Bragg peaks still can be observed at $Q$ = (0, 0, $n$ + 1/2) at $T$ = 280 K [Figs.~\ref{fig1}(i) and \ref{fig1}(k)]. A similar phenomenon has been reported in Ref.~\cite{Meier2019} and was explained as the tails of inelastic scattering by low-energy transverse magnons. In our study, we can rule out this possibility in Fe$_{0.89}$Co$_{0.11}$Sn clearly (see details in Sec. IIIB) and confirm the existence of the small in-plane magnetic moment components which result in the weak magnetic peaks at $Q$ = (0, 0, $n$ + 1/2) in the axial AFM state. At last, we calculate the component of the ordered moment along the $c$ axis, $m_c\approx1.39~\mu_{B}$, and the small in-plane component $m_{ab}\approx0.12~\mu_{B}$, which corresponds to a small canting angle ($\sim$4.84$\degree$) of the ordered spins away from the $c$ axis (see details in the SM~\cite{SI}).

\begin{figure*}[htb]
\center{\includegraphics[width=0.9\linewidth]{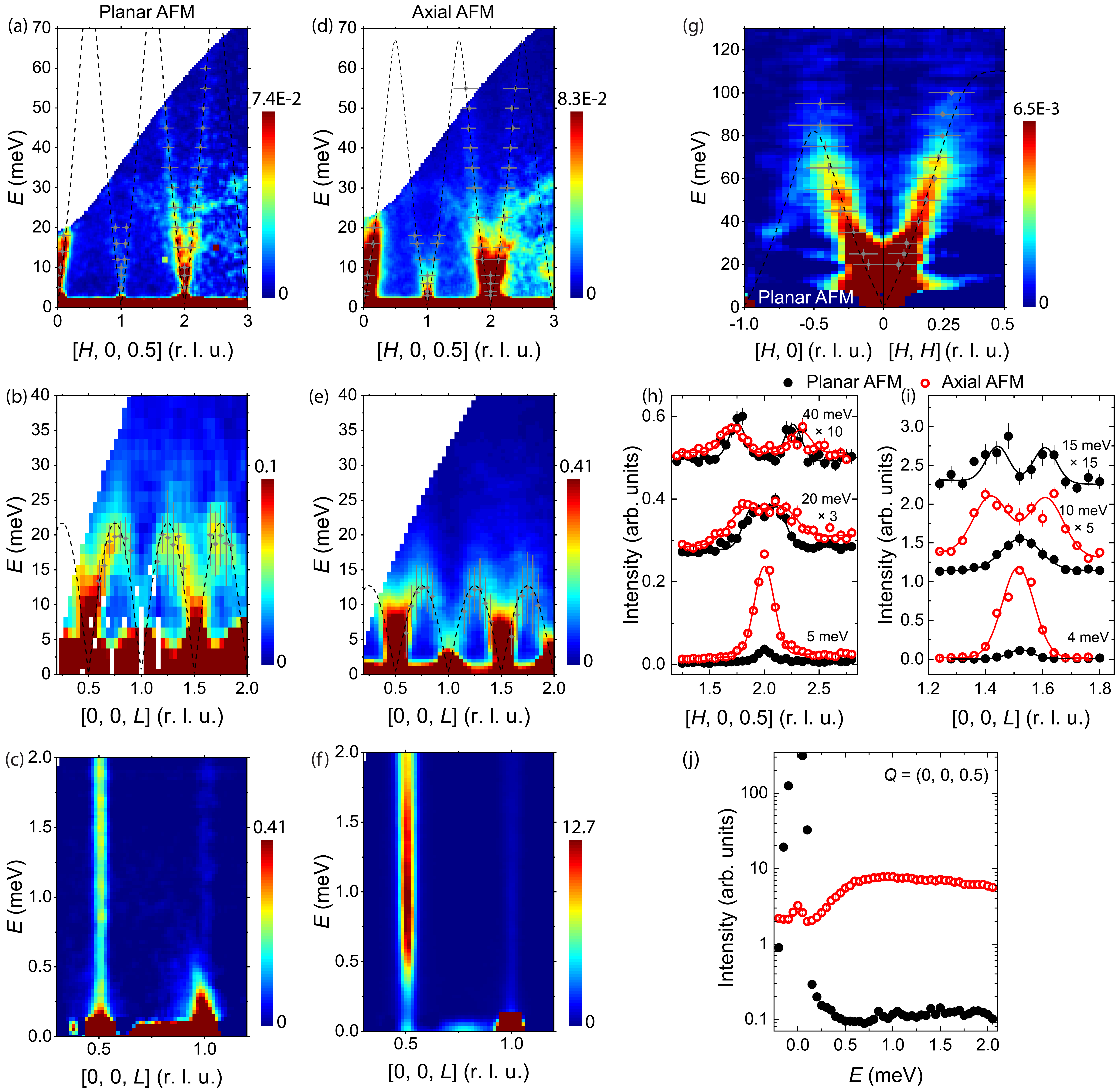}}
  \caption{~Dispersion of the spin wave. (a) and (b) Spin-wave dispersion along the [$H$, 0, 0.5] direction [the red path in Fig.~\ref{fig1}(d)] measured with $E_i$ = 80 meV, $T$ = 10 K, and along the [0, 0, $L$] direction measured with $E_i$ = 150 meV, $T$ = 10 K. (c) Spin-wave dispersion along the [0, 0, $L$] direction measured with $E_i$ = 3.32 meV, $T$ = 10 K. (d) and (e) Spin-wave dispersion along the [$H$, 0, 0.5] and [0, 0, $L$] directions, respectively, measured with $E_i$ = 80 meV, $T$ = 280 K. (f) Spin-wave dispersion along the [0, 0, $L$] direction measured with $E_i$ = 3.32 meV, $T$ = 280 K. The extra intensities in (a) and (d) around 10 and 30 meV in the high-$Q$ region are the phonon signal. (g) Spin-wave dispersion along the [$H$, 0]$\rightarrow$[$H$, $H$] path at 10 K measured with $E_i$ = 150 meV. The extra intensity around 10 meV is the intensity of phonon scattering. (h) and (i) 1D constant-energy curves along the [$H$, 0, 0.5] direction and the [0, 0, $L$] direction at different energies in both planar and axial AFM states. The solid lines are fittings with the Gaussian function. (j) 1D constant-momentum cuts at $Q$ = (0, 0, 0.5) in both planar and axial AFM states. The integration range along the [$-K$/2 $K$ 0] direction for all the panels is $K$ = [$-$0.1, 0.1]. The dashed lines in (a), (b), (d), (e), and (g) are results of LSWT fittings (see details in Sec. IV). The data points with error bars in (a), (d), and (g) are the peak positions of the 1D constant-energy curves shown in (h) and (i). The vertical error bars are the energy resolution of the instrument, and the horizontal error bars are the full width at half maximum (FWHM) of the Gaussian fittings there. The data points in (b) and (e) are peak positions of the 1D constant-momentum curves. The vertical error bars are the line width (FWHM) of spectra, and the horizontal error bars are the integration momentum range. }
\label{dispersion_low}
\end{figure*}

\subsection{Low- and intermediate-energy spin waves}
%\subsubsection{Planar AFM state}
Now the spin dynamics of Fe$_{0.89}$Co$_{0.11}$Sn are discussed. Figure~\ref{dispersion_low} presents the spin-wave results in the low and intermediate energy range, which were collected with $E_i$ = 3.32 meV (at CNCS), $E_i$ = 80 meV (at SEQUOIA), and $E_i$ = 150 meV (at ARCS, only $T$ = 10 K). There are three magnetic (Fe or Co) atoms in one unit cell [Figs.~\ref{fig1}(b) and ~\ref{fig1}(c)], which will give rise to three magnon branches. As shown in Figs.~\ref{dispersion_low}(a) and \ref{dispersion_low}(d), we can see the steep dispersion of the acoustic magnon along the [$H$, 0, 0.5] direction for both planar (10 K) and axial (280 K) AFM states. While the top of the acoustic spin-wave band of the planar AFM state cannot be clearly seen in the measured energy range with $E_i$ = 80 meV, the energy band top for the axial AFM state seems to appear at $\sim$67 meV. Figure~\ref{dispersion_low}(g) shows spin-wave dispersion along the [$H$, 0]$\rightarrow$[$H$, $H$] path at 10 K measured with $E_i$ = 150 meV. The top of the acoustic spin-wave band at the $M$ point appears around 82 meV, above which the weak spin-excitation intensity continues up to 130 meV. On the other hand, the spin wave dispersion along the [0, 0, $L$] direction reaches the band top at about 21 and 13 meV for the planar and axial AFM states, respectively [Figs.~\ref{dispersion_low}(b) and \ref{dispersion_low}(e)]. We extracted 1D constant-energy curves from the spin-wave dispersion in Figs.~\ref{dispersion_low}(a), \ref{dispersion_low}(b),~\ref{dispersion_low}(d), and \ref{dispersion_low}(e), and fitted the curves with Gaussian functions. Some of the 1D constant-energy curves are shown in Figs.~\ref{dispersion_low}(h) and \ref{dispersion_low}(i). The data points in Figs.~\ref{dispersion_low}(a), \ref{dispersion_low}(d), and \ref{dispersion_low}(g) are the peak positions of these 1D constant-energy curves.

\begin{figure*}[t]
\center{\includegraphics[width=0.9\linewidth]{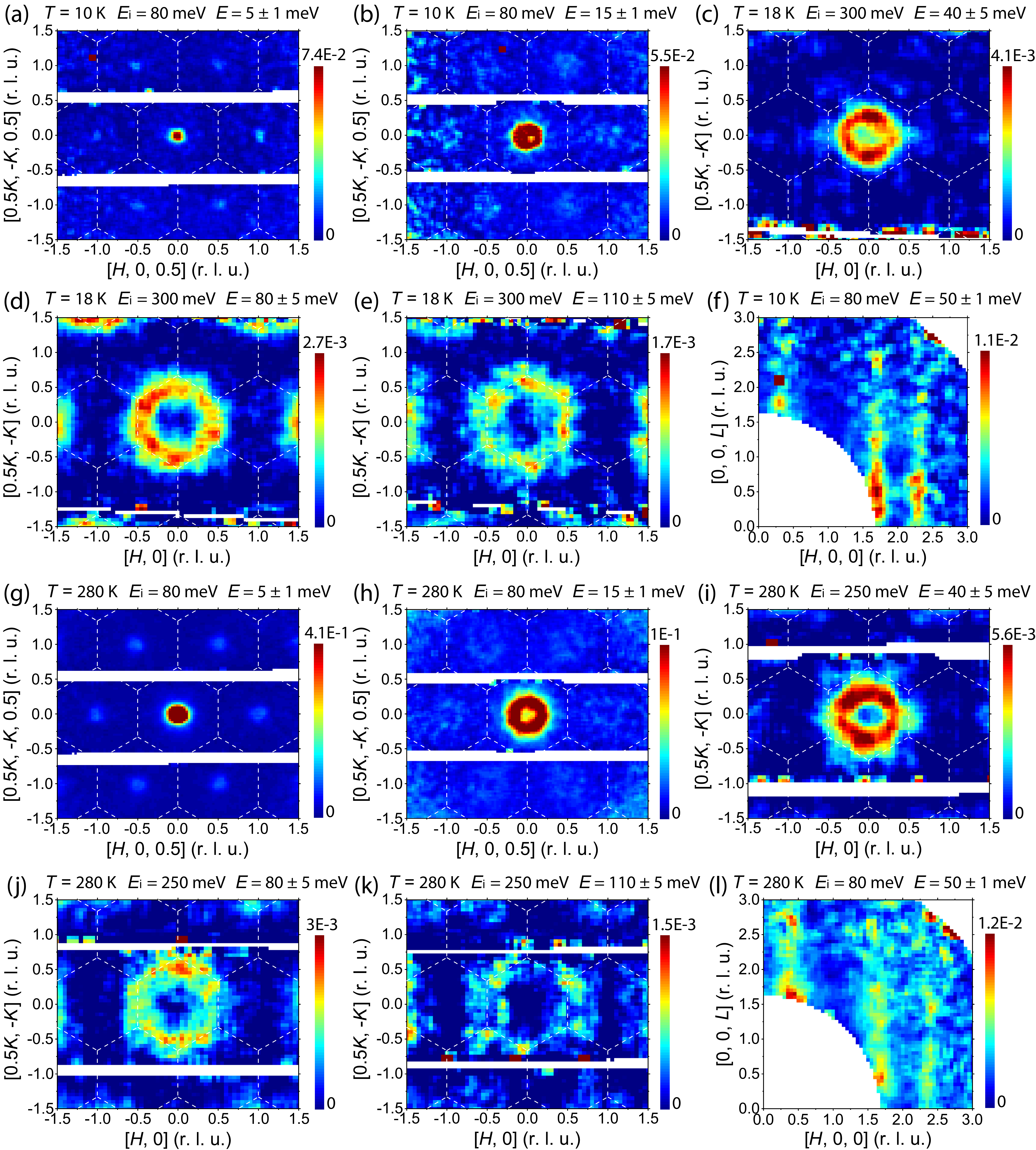}}
  \caption{~Constant-energy 2D slices in the ($H$ $K$) plane and the ($H$ 0 $L$) plane. (a)--(e) The evolution of the spin wave with the increasing energy in the ($H$ $K$) plane at $T$ = 10 and 18 K (planar AFM state). (f) Constant-energy slice in the ($H$ 0 $L$) plane at $T$ = 10 K and $E$ = 50 $\pm$ 1 meV. (g)--(k) The evolution of the spin wave with the increasing energy in the ($H$ $K$) plane at $T$ = 280 K (axial AFM state). (l) Constant-energy slice in the ($H$ 0 $L$) plane at $T$ = 280 K and $E$ = 50 $\pm$ 1 meV. The white dashed lines represent the boundary of the Brillouin zones.
  }
  \label{constantE}
\end{figure*}

\begin{figure*}[tb]
\center{\includegraphics[width=1\linewidth]{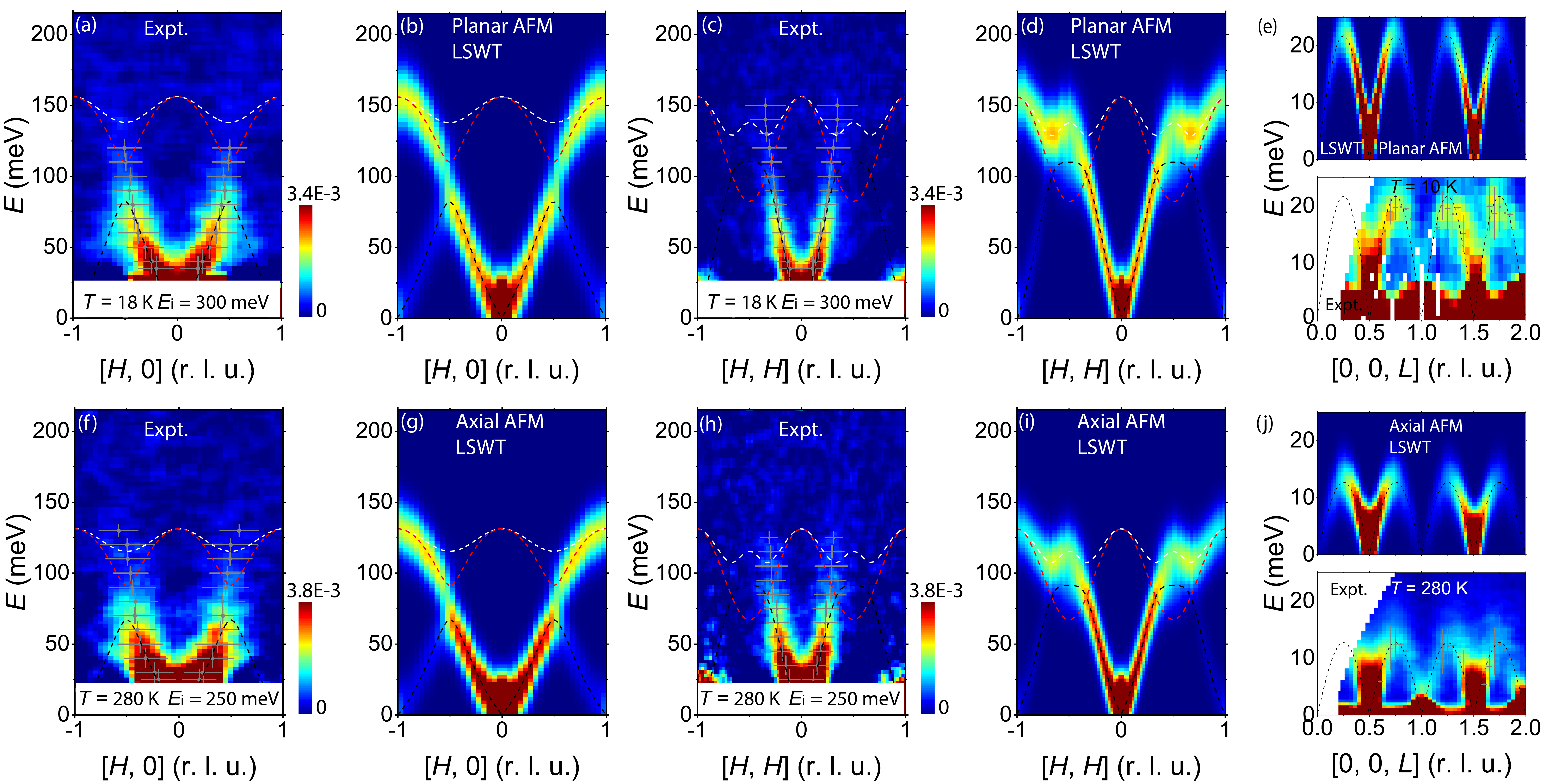}}
  \caption{~The high-energy spin wave and LSWT fittings with SPINW. (a)--(d) Experimental INS spectra along the [$H$, 0] [the red path in Fig.~\ref{fig1}(d)] and [$H$, $H$] [the green path in Fig.~\ref{fig1}(d)] directions measured with $E_i$ = 300 meV, $T$ = 18 K and the corresponding LSWT calculations. (e) Comparison of the experimental spin-wave dispersion and the LSWT calculation of the spin-wave dispersion along [0, 0, $L$] for the planar AFM state.
  (f)--(i) INS spectra along the [$H$, 0] and [$H$, $H$] directions measured with $E_i$ = 250 meV, $T$ = 280 K and the corresponding LSWT calculations. (j) Comparison of the experimental spin-wave dispersion and the LSWT calculation of the spin-wave dispersion along [0, 0, $L$] for the axial AFM state. Some extra intensities that are away from the main dispersion around 50 meV in (a), (c), (f), and (h) are residual intensities due to the imperfect background subtraction. The gray data points with error bars in (a), (c), (f), and (h) are fitted peak positions of constant-energy curves. The vertical error bars are the energy resolution of the instrument, and the horizontal error bars are the FWHM of the Gaussian fittings. The dashed lines are LSWT calculations with the best fitting parameters in Table~\ref{spinwparameter}. The black dashed lines indicate the acoustic magnons, and the red and white dashed lines indicate the optical magnons.
  }
  \label{dispersion_high}
\end{figure*}

In order to figure out whether the spin wave of Fe$_{0.89}$Co$_{0.11}$Sn is gapped or gapless, we measured the low-energy excitations with $E_i$ = 3.32 meV (with an energy resolution $\sim$0.11 meV at the elastic position~\cite{SI}). From Figs.~\ref{dispersion_low}(c) and \ref{dispersion_low}(f), we can see sharp spin waves stem from $Q$ = (0, 0, 0.5) for both states. The 1D energy cuts at $Q$ = (0, 0, 0.5) show the evolution of the spin-wave intensity more intuitively [Fig.~\ref{dispersion_low}(j)]. At 10 K, we do not see the abrupt intensity decrease with the decreasing energy at $Q$ = (0, 0, 0.5), which was observed and considered as evidence of the spin gap in FeSn~\cite{FeSn2021flat}. This indicates that the spin wave of the planar AFM state in Fe$_{0.89}$Co$_{0.11}$Sn is gapless in our resolution limit. At 280 K, the intensity of the spin wave first gradually increases with the decreasing energy above 1 meV, then abruptly decreases with the decreasing energy below $\sim$0.5 meV until the spin wave touches the tail of the Bragg peak, and upturns below $\sim$0.1 meV [Fig.~\ref{dispersion_low}(j)]. Having this high-resolution data at the axial AFM state, two important issues can be figured out clearly. First, in the axial AFM state, there is a small spin gap below $\sim$0.5 meV, although the gap is not fully opened. Second, the weak magnetic Bragg peaks observed at $Q$ = (0, 0, $n$ + 1/2) (see Sec. IIIA) are intrinsic magnetic peaks for the existence of in-plane magnetic moments in the axial AFM state. These peaks cannot be the tails of inelastic scattering by low-energy transverse magnons~\cite{Meier2019} because the inelastic tail should have a rather low intensity below $\sim$0.2 meV for the opening of the spin gap. If the elastic peaks at $Q$ = (0, 0, $n$ + 1/2) are inelastic tails, their intensity should be lower than (or comparable to) the intensity of the excitations at $\sim$0.2 meV. With this, there cannot be the obviously upward intensity below $\sim$0.1 meV as shown in Fig.~\ref{dispersion_low}(j). Thus, we demonstrate that the weak magnetic Bragg peaks observed at $Q$ = (0, 0, $n$ + 1/2) at 280 K are intrinsic, and the magnetic moments are not perfectly aligned along the $c$ axis in the axial AFM state of Fe$_{0.89}$Co$_{0.11}$Sn.

\subsection{High-energy spin excitations}

To cover the high-energy spin excitations of Fe$_{0.89}$Co$_{0.11}$Sn, we measured the spin dynamics with higher incident neutron energies: $E_i$ = 300 meV (at ARCS and $T$ = 18 K) and 250 meV (at SEQUOIA and $T$ = 280 K). Figure~\ref{constantE} presents some constant-energy 2D slices in the ($H$ $K$) plane and the ($H$ 0 $L$) plane. In the ($H$ $K$ 0.5) plane, the spin waves stem from the same positions where the magnetic Bragg peaks are observed in Figs.~\ref{fig1}(j) and \ref{fig1}(k). The small spots then evolve to be circles with increasing energy. From Figs.~\ref{constantE}(a)--\ref{constantE}(c) and \ref{constantE}(g)--\ref{constantE}(i), we can see that the sizes of the spots or circles in the axial AFM state (280 K) are always larger than those in the planar AFM state (10 or 18 K), which means that the spin excitation can reach the BZ boundary at a lower energy in the axial AFM state and is consistent with the analysis in Sec. IIIB. When the neutron energy transfer approaches $\sim$80 meV, the spin excitations evolve to the edge of the BZs, which indicates the energy top of the acoustic magnon band. Furthermore, we found that the spin excitation has no obvious intensity modulation along the [0, 0, $L$] direction for energies above $\sim$30 meV~[Figs.~\ref{constantE}(f) and \ref{constantE}(l)]. Thus the results above 30 meV shown in Figs.~\ref{constantE} and~\ref{dispersion_high} were extracted by integrating a wide range of $L$ ($-$5 $\le$ $L$ $\le$ 5), and we will omit the $L$ indices for these cases.

The INS spectra up to 215 meV are shown in Fig.~\ref{dispersion_high}. For the planar AFM state, the dispersion along the [$H$, 0] direction shows strong intensity below  $\sim$100 meV, above which the signals become diffusive and rather weak, but still can be identified up to $\sim$200 meV [Fig.~\ref{dispersion_high}(a)]. For the dispersion along the [$H$, $H$] direction [Fig.~\ref{dispersion_high}(c)], a sharp spin wave below $\sim$100 meV and an obvious intensity decrease above $\sim$100 meV are also observed. In the axial AFM state [Figs.~\ref{dispersion_high}(f) and~\ref{dispersion_high}(h)], the high-energy spin excitations are similar to those in the planar AFM state, while the energy scale is smaller than in the planar AFM state. It is worth noting that here we only can observe the clear acoustic spin-wave mode and the weak diffusive spin excitations (between $\sim$90 and $\sim$200 meV); any indications of the other two expected magnon modes cannot be identified.
\renewcommand{\arraystretch}{1.3}
\begin{table*}[thb!]
\caption {Exchange coupling parameters (meV) for the magnetic Hamiltonian~(\ref{Hamiltonian}) obtained from the LSWT fittings.}
 \begin{ruledtabular}
%  \begin{tabular}{l  l  l  l  l  l}
 \begin{tabular}{cccccc}
 Ordering type    &   $SJ_1$    &   $SJ_2$     &    $SJ_c$    &  $SK_c$
 &  $SK_a$ \\
 \hline
 Planar AFM       &   -18.15~$\pm$~4.9    &   -4.50~$\pm$~1.89      &    10.87~$\pm$~2.21        &   0~$ < K_c \le$~0.0038 &   \\
 %\hline
 Axial AFM        &   -15.90~$\pm$~4.32    &   -3.95~$\pm$~1.55       &    6.37~$\pm$~1.54     &   -0.009 &   -0.009 \\
 \end{tabular}
\end{ruledtabular}
\label{spinwparameter}
\end{table*}
\section{LSWT simulations and discussions}
To understand the experimentally observed magnetic excitations, we employ the LSWT simulations using the SPINW library~\cite{SpinW}. We use the following Heisenberg Hamiltonian:
\begin{align}  \label{Hamiltonian}
	&\mathcal{H} = J_1\sum_{\langle i,j \rangle} \mathbf{S}_i \cdot \mathbf{S}_j + J_2\sum_{\langle\langle i,j \rangle\rangle} \mathbf{S}_i \cdot \mathbf{S}_j + J_c\sum_{\langle i,j \rangle} \mathbf{S}_i \cdot \mathbf{S}_j \nonumber \\
	&+K_c\sum_{\langle i \rangle} (S^z_i)^2 +  K_a\sum_{\langle i \rangle} (S^{{\hat e}_{\mathbf{r}}}_i)^2,
\end{align}
where $J_1$ is the in-plane nearest-neighbor (NN) exchange coupling, $J_2$ is the in-plane next-nearest-neighbor (NNN) exchange coupling, and $J_c$ is the NN interlayer exchange coupling. The last two terms represent the single-ion anisotropy. It should be noted that the last term is only applicable for the axial AFM state. With $K_a < 0$, $K_a(S^{{\hat e}_{\mathbf{r}}}_i)^2$ represents the in-plane easy-axis anisotropy following the lattice symmetry, which is responsible for the small canting angle of the ordered spins. The in-plane easy-axis direction ${\hat e}_{\mathbf{r}}$ depends on the position ($\mathbf{r}$) of the magnetic atoms (see details in the SM~\cite{SI}). Due to the itinerant properties of such a metallic system, the spin value could be ambiguous. The effective spin value may also change from the planar AFM state to the axial AFM state. To describe the magnetic Hamiltonian smoothly, we thus use the combination of the spin value and exchange coupling parameters $SJ_1$, $SJ_2$, $SJ_c$, $SK_c$, and $SK_a$ hereinafter.

We first cut the experimental spectra and fitted them to get the dispersion relation and intensity of the spin wave. Then we fit the extracted data using SPINW with Hamiltonian~(\ref{Hamiltonian}) to get the exchange coupling parameters (see details in the SM~\cite{SI}). The best fitting parameters are summarized in Table~\ref{spinwparameter}. The fitted spin-wave dispersion curves for the acoustic magnon are shown as dashed lines in Figs.~\ref{dispersion_low}(a), \ref{dispersion_low}(b), \ref{dispersion_low}(d), \ref{dispersion_low}(e), and~\ref{dispersion_low}(g), which indeed can describe the data points from the experimental results perfectly. The calculated dynamical spin structure factors with the parameters in Table~\ref{spinwparameter} are shown in Figs.~\ref{dispersion_high}(b), \ref{dispersion_high}(d), \ref{dispersion_high}(e), \ref{dispersion_high}(g), \ref{dispersion_high}(i) and \ref{dispersion_high}(j).

Note that the single-ion anisotropy parameters $SK_c$ and $SK_a$ in Table~\ref{spinwparameter} are determined separately from $SJ_1$, $SJ_2$, and $SJ_c$ ~\cite{SI}. This is a reasonable approach since the anisotropy terms are rather small and have negligible influence on the spin-wave dispersion here. Although no spin gap can be identified in the planar AFM state, it is still necessary to include an easy-plane single-ion anisotropy ($SK_c > 0$) to confine the ordered spins in the $ab$ plane. As for the axial AFM state, a minor spin gap below $~0.5$ meV has been identified, which requires a nonzero single-ion anisotropy to open the gap. However, as the small canting angle ($\sim$4.84$\degree$) has been confirmed by our previous analysis in Secs. IIIA and IIIB, a single easy-axis anisotropy term ($SK_c < 0$) cannot stabilize such a special magnetic structure. Our solution is to add the $K_a\sum_{\langle i \rangle} (S^{{\hat e}_{\mathbf{r}}}_i)^2$ term ($SK_a < 0$) in the Hamiltonian. At last, we estimate the single-ion anisotropy parameters to be $0 < SK_c \le$~0.0038 meV in the planar AFM state and $SK_c$ = $SK_a$ = $-$0.009 meV in the axial AFM state (see details in the SM~\cite{SI}).

From Table~\ref{spinwparameter}, we can see that the generalized exchange coupling parameters $SJ_1$, $SJ_2$, and $SJ_c$ decrease on different levels from the planar AFM state to the axial AFM state. Specifically, ($SJ_1$)$_{\text{axial}}/$($SJ_1$)$_{\text{planar}}\approx$ 0.876, ($SJ_2$)$_{\text{axial}}/$($SJ_2$)$_{\text{planar}}\approx$ 0.878,
($SJ_c$)$_{\text{axial}}/$($SJ_c$)$_{\text{planar}}\approx$ 0.586. If the effective local spin value in the Hamiltonian is supposed to be proportional to the ordered magnetic moment of the ground state, we can get the ratio of the effective spin value between the planar and the axial AFM states: $S^{\text{axial}}/S^{\text{\text{planar}}}$ = $m^{\text{axial}}/m^{\text{\text{planar}}}\approx$ 0.760 (see details in the SM~\cite{SI}). Then the ratios of the real exchange coupling parameters: $J^{\text{axial}}_1/J^{\text{planar}}_1\approx$ 1.153, $J^{\text{axial}}_2/J^{\text{planar}}_2\approx$ 1.155, $J^{\text{axial}}_c/J^{\text{planar}}_c\approx$ 0.771. This means that from the planar AFM state to the axial AFM state, the in-plane parameters $J_1$ and $J_2$ increase synchronously, while the out-of-plane parameter $J_c$ decreases. These interesting evolutions of the effective exchange couplings should reflect the changes in the electronic band structure for the planar-to-axial phase transition, which have been demonstrated by theoretical calculations of the electronic structure on a similar compound, Fe$_{0.94}$Co$_{0.06}$Sn~\cite{Moore2022}. To explain the observed changes in the effective exchange coupling parameters on the basis of the change in the electronic structure, one should project the exchange coupling interactions into orbital-resolved contributions. Exchange coupling parameters can be calculated from the electronic structure using the known formalisms, e.g., the local spin density functional~\cite{Liechtenstein1987} or the real-space linear-muffin-tin method~\cite{Schilfgaarde1999,Frota2000}, which require extensive computational work that is out of the scope of this paper.
\begin{figure*}[thb!]
\center{\includegraphics[width=0.7\linewidth]{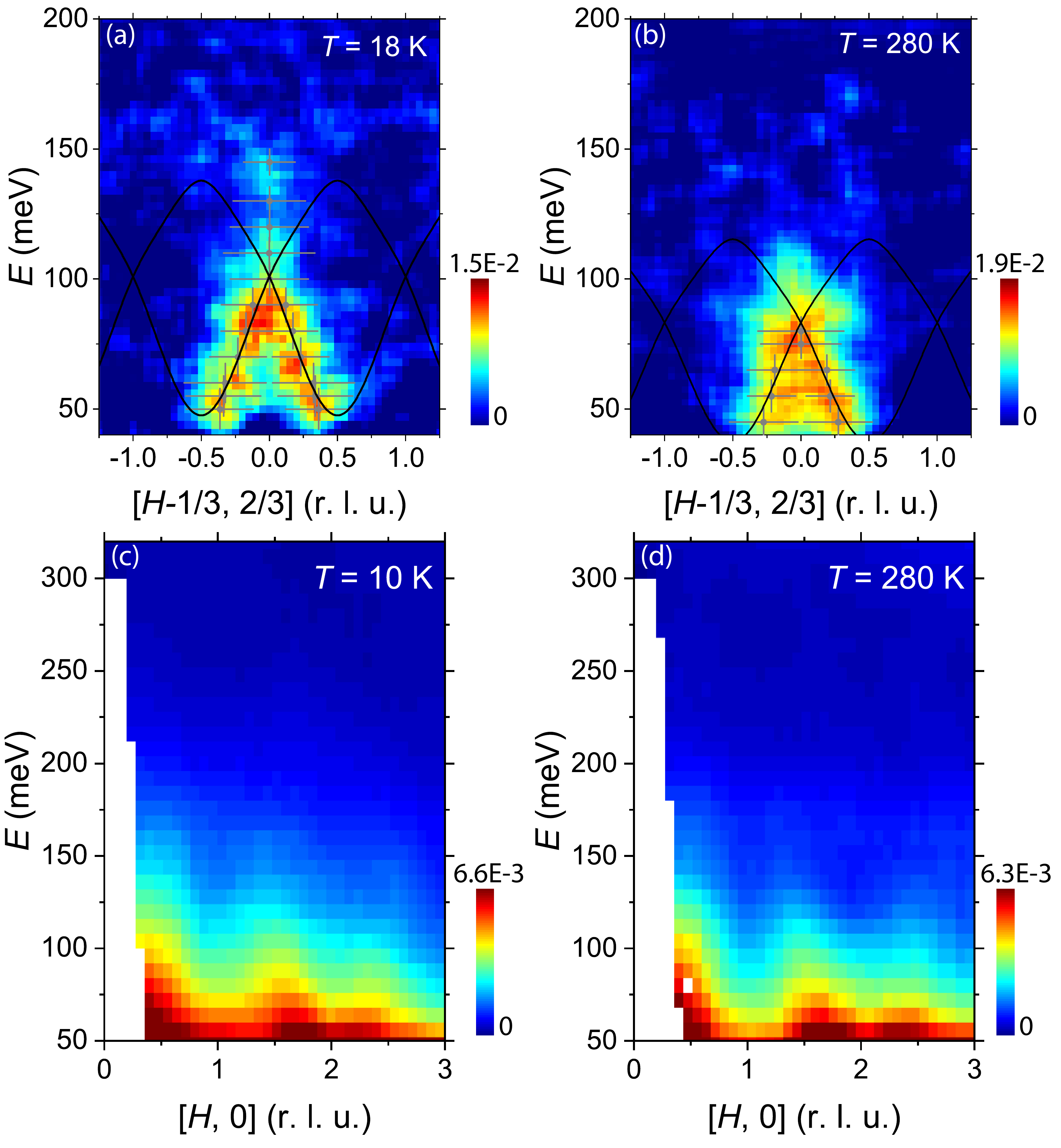}}
  \caption{~(a) and (b) Spin-excitation dispersion along the [$H-$1/3, 2/3] direction [the blue path in Fig.~\ref{fig1}(d)] in planar (a) and axial (b) AFM states. The gray data points with error bars are fitted peak positions of constant-energy curves. The vertical error bars are the energy resolution of the instrument, and the horizontal error bars are the FWHM of the Gaussian fittings. The black solid lines are the calculated spin-wave dispersion using the LSWT with the parameters in Table~\ref{spinwparameter}. (c) and (d) High-energy spin excitations measured with $E_i$ = 400 meV.
  }
  \label{dirac}
\end{figure*}

By comparing the experimental and LSWT simulation results, it is clear that the LSWT calculation works well only for the acoustic spin wave. For the spin excitations above the acoustic magnon, the data look ambiguous, and we cannot identify the residual two optical magnon modes. The LSWT simulation cannot cover the experimental results. The weak and ambiguous excitations above the acoustic magnon and the disappearance of the optical magnon modes can be explained by the interaction between the spin wave and the Stoner excitations from the itinerant magnetism~\cite{Mook1971spin,Mook1974magnetic,Ishikawa1078,Romanov1988,melnikov2018,Chen2020unconventional,Coey2021handbook,FeSn2021flat,do2021damped,Zhang2022FeSncalculation}.
As depicted in Fig.~\ref{fig1}(e), we only can observe well-defined spin wave at the relatively low-energy region (below $\sim$90 meV here) just before touching the lower boundary of the Stoner continuum. After entering the Stoner continuum (above $\sim$90--100 meV here), the spin waves decay into the particle-hole excitations, which makes the optical magnon modes invisible and only leaves us the observable weak damped excitations up to $\sim$250 meV. A recent $ab~initio$ study on FeSn indicates that the Stoner continuum appears above $\sim$~80--100 meV and overlaps with the high-energy magnon spectra, which results in the strong damping of the magnon~\cite{Zhang2022FeSncalculation}. Our results in Fe$_{0.89}$Co$_{0.11}$Sn here are qualitatively consistent with this calculation and the experimental INS results in FeSn~\cite{do2021damped}.

We further check the data across the $K$ points of BZs [the blue path shown in Fig.~\ref{fig1}(d)] in both AFM states. In Fig.~\ref{dirac}(a), we can see that the spectrum shows a downward-cone-like shape with the vertex appearing at $\sim$100 meV at $T$ = 18 K. Above the downward cone, the excitation intensity becomes weak and diffusive and is similar to the aforementioned results shown along the red and green paths in Fig.~\ref{fig1}(d). Similarly, at $T$ = 280 K, a downward cone with a slightly lower vertex ($\sim$85 meV) can also be identified [Fig.~\ref{dirac}(b)], although it is not as clear as that at $T$ = 18 K. This feature is not easy to understand at first glance. However, if we consider the aforementioned interaction between the spin wave and the itinerant Stoner continuum, the downward-cone-like excitation here should be the lower part of two crossed spin-wave branches, with the upper part of the crossed branches becoming weak and diffusive from entering the so-called Stoner continuum. The band-crossing-like features can be further supported by the LSWT.  In Figs.~\ref{dirac}(a) and~\ref{dirac}(b), the black solid lines are the calculations from the LSWT using the parameters in Table~\ref{spinwparameter}. The calculations indeed show band crossings at the vertices of the measured downward cones. Such a kind of band crossing is known as the criterion of Dirac magnons~\cite{Yao2018topological,Bao2018discovery,Yuan2020dirac,Elliot2021order,mcclarty2021topological}. This indicates that we may have found experimental evidence for the existence of Dirac magnons in Fe$_{0.89}$Co$_{0.11}$Sn. However, the existence of the itinerant Stoner continuum stops us from investigating this feature further. We note that a similar band crossing feature was also observed in pure FeSn and argued to be damped Dirac magnons~\cite{do2021damped}.

Another noteworthy point is that a magnetic flat band from the quasiparticle excitations between the spin-up flat electronic band (majority electrons) and spin-down flat electronic band (minority electrons) has been proposed for 2D FM metals with a kagome lattice~\cite{FeSn2021flat}. In the case of FeSn, despite the presence of AFM order below $T_N$, it was treated as a quasi-2D FM metal for the weak AFM coupling between the adjacent FM planes~\cite{FeSn2021flat,do2021damped}. Starting from such a quasi-2D FM metal, the theoretical calculations predicted a magnetic flat band of the spin excitations in FeSn. However, according to the INS results such a flat band is absent up to $\sim$300 meV in FeSn~\cite{FeSn2021flat,do2021damped}. Compared with the pure FeSn, Fe$_{0.89}$Co$_{0.11}$Sn has a similar AFM transition temperature ($T_N$~$\approx$~340 K), the same planar AFM order (below $T_{N}^{''}$~$\approx$~70 K), and comparable ordered magnetic moment~\cite{SI}.
Therefore we would expect that Fe$_{0.89}$Co$_{0.11}$Sn has a similar electronic structure and itineracy to FeSn. This means that the predicted magnetic flat band for FeSn~\cite{FeSn2021flat} is expected to exist in Fe$_{0.89}$Co$_{0.11}$Sn. Since such a flat band has not been observed up to $\sim$210 meV (measurements with $E_i$ = 300 meV) in our Fe$_{0.89}$Co$_{0.11}$Sn sample, we then measured the higher-energy spin excitations with $E_i$ = 400 meV to see if there is a magnetic flat band in higher-energy regions. We found the strong spin-wave dispersion below $\sim$100 meV and the weak Stoner continuum intensity up to $\sim$250 meV only [Figs.~\ref{dirac}(c) and \ref{dirac}(d)]. Our results show that there is no sign of the localized magnetic flat band in Fe$_{0.89}$Co$_{0.11}$Sn up to $\sim$320 meV. The absence of such a flat band in experiments could have three possible reasons: (i) The flat band is too weak to be visible. (ii) The flat band may mix with the general Stoner continuum from other transition channels and could lose its flatness and narrowness~\cite{Zhang2022FeSncalculation}. Together with the possibly low intensity as mentioned in reason (i), the flat band could become indistinguishable. (iii) The flat band does not exist, which is inconsistent with the theoretical prediction~\cite{FeSn2021flat}.

\section{Summary}
We have performed systematic neutron scattering measurements on the kagome-lattice AFM metal Fe$_{0.89}$Co$_{0.11}$Sn. The planar and axial AFM ordered states are confirmed by neutron diffraction and magnetization results. The careful analyses of the diffraction and the low-energy spin-wave results demonstrate that the weak magnetic Bragg peaks at $Q$ = (0, 0, $n$ + 1/2) ($n$ = integer) of the axial state are intrinsic and come from the small in-plane magnetic moment components. Although it has been well confirmed that the ordered moments stack antiferromagnetically along the $c$ axis, the spin-excitation spectra are dominated by the in-plane FM spin excitation, which indicates quasi-2D magnetism. The INS shows a sharp spin wave below $\sim$90 meV, above which the spin excitations become weak and diffusive, but persist up to $\sim$250 meV. The sharp acoustic spin-wave band can be described in the frame of LSWT by a Heisenberg $J_1$-$J_2$-$J_c$ model considering weak single-ion anisotropy. In the axial AFM state, although the generalized exchange coupling parameters $SJ_1$, $SJ_2$, and $SJ_c$ are smaller than those in the planar state, $J_1$ and $J_2$ may show the opposite behavior if the change in effective spin value is considered. Above the acoustic magnon, the Stoner continuum appears, which makes the optical magnons highly damped and invisible. At the $K$ points of the BZs, we give evidence for the existence of the Dirac magnon with the upper part of the Dirac cone becoming weak and decayed in the Stoner continuum. The magnetic flat band is demonstrated to be absent in Fe$_{0.89}$Co$_{0.11}$Sn up to $\sim$320 meV. Our results give a comprehensive overview of the INS experiments and LSWT calculations on kagome-lattice AFM metal Fe$_{0.89}$Co$_{0.11}$Sn. The absence of the two optical magnon branches and the upper part of the Dirac cone highlights the indispensable role of the itinerant electrons in understanding the magnetism in Fe$_{0.89}$Co$_{0.11}$Sn.

\
\section*{Acknowledgments}
We thank Dr. Victor Fanelli for help with the experiment at the SEQUOIA spectrometer, Dr. Jong Keum for help with the x-ray Laue measurements, and Dr. Matthew Stone for suggestions about the neutron beam time application. Work at Oak Ridge National Laboratory (ORNL) was supported by the U.S. Department of Energy (DOE), Office of Science, Basic Energy Sciences, Materials Science and Engineering Division. H.L. was supported by the National Key R$\&$D Program of China (Grant No. 2018YFE0202600, 2022YFA1403800), the Beijing Natural Science Foundation (Grant No. Z200005), and the National Natural Science Foundation of China (12274459).  This research used resources at the Spallation Neutron Source, a DOE Office of Science User Facility operated by the Oak Ridge National Laboratory.
X-ray Laue measurements were conducted at the Center for Nanophase Materials Sciences (CNMS) (CNMS2019-R18) at ORNL, which is a DOE Office of Science User Facility.

\end{document}